\documentclass[sigconf]{acmart}
\copyrightyear{2020}
\acmYear{2020}
\setcopyright{acmcopyright}
\acmConference[CIKM '20] {The 29th ACM International Conference on Information and Knowledge Management}{October 19--23, 2020}{Virtual Event, Ireland}
\acmBooktitle{The 29th ACM International Conference on Information and Knowledge Management (CIKM '20), October 19--23, 2020, Virtual Event, Ireland}
\acmPrice{15.00}
\acmDOI{10.1145/3340531.3411996}
\acmISBN{978-1-4503-6859-9/20/10}
\usepackage{booktabs}
\usepackage{multirow}
\usepackage{subcaption}
\usepackage[ruled,linesnumbered]{algorithm2e}
\usepackage{enumitem}
\newcommand{\tabincell}[2]{\begin{tabular}{@{}#1@{}}#2\end{tabular}}
\begin{document}
\fancyhead{}
\title{DisenHAN: Disentangled Heterogeneous Graph Attention Network for Recommendation}

\author{Yifan Wang}
\affiliation{%
  \institution{Department of Computer Science, School of EECS, Peking University}
}
\email{yifanwang@pku.edu.cn}

\author{Suyao Tang}
\affiliation{%
   \institution{Department of Computer Science, School of EECS, Peking University}
}
\email{tangsuyao@pku.edu.cn}

\author{Yuntong Lei}
\affiliation{%
   \institution{Department of Computer Science, School of EECS, Peking University}
}
\email{leiyunton@pku.edu.cn}

\author{Weiping Song}
\affiliation{%
   \institution{Department of Computer Science, School of EECS, Peking University}
}
\email{songweiping@pku.edu.cn}

\author{Sheng Wang}
\affiliation{%
   \institution{Paul G. Allen School of Computer Science, University of Washington}
}
\email{swang@cs.washington.edu}

\author{Ming Zhang}
\authornote{Ming Zhang is the corresponding author.}
\affiliation{%
   \institution{Department of Computer Science, School of EECS, Peking University}
}
\email{mzhang_cs@pku.edu.cn}

\begin{abstract}
Heterogeneous information network has been widely used to alleviate sparsity and cold start problems in recommender systems since it can model rich context information in user-item interactions. Graph neural network is able to encode this rich context information through propagation on the graph. However, existing heterogeneous graph neural networks neglect entanglement of the latent factors stemming from different aspects. Moreover, meta paths in existing approaches are simplified as connecting paths or side information between node pairs, overlooking the rich semantic information in the paths. In this paper, we propose a novel disentangled heterogeneous graph attention network DisenHAN for top-$N$ recommendation, which learns disentangled user/item representations from different aspects in a heterogeneous information network.
In particular, we use meta relations to decompose high-order connectivity between node pairs and propose a disentangled embedding propagation layer which can iteratively identify the major aspect of meta relations. Our model aggregates corresponding aspect features from each meta relation for the target user/item. With different layers of embedding propagation, DisenHAN is able to explicitly capture the collaborative filtering effect semantically. Extensive experiments on three real-world datasets show that DisenHAN consistently outperforms state-of-the-art approaches. We further demonstrate the effectiveness and interpretability of the learned disentangled representations via insightful case studies and visualization.
\end{abstract}






\maketitle



\section{Introduction}
Recommender systems play an important role for guiding a user in a personalized way of discovering interested products from overwhelming alternatives. Collaborative filtering (CF) based methods (e.g., matrix factorization\cite{koren2009matrix}) have been extensively used for recommendation, assuming users who have made similar choices tend to have similar preferences in the future. Since CF based methods usually suffer from the sparsity of user-item interactions and the cold start problems, many methods integrate other context information by transforming the interaction into a feature-rich data instance for user and item. These instances are then used by supervised learning models to predict the score\cite{cheng2016wide, song2019autoint}. However, these methods always treat each user-item record as an independent instance and overlook the relations among 
them.

Heterogeneous information network (HIN), which consists of multiple types of nodes and/or links, is flexible to model the heterogeneity and complexity of user-item interactions with rich context information. In particular, meta path, a composite relation connecting node pairs in HIN, is widely used to capture the relevant semantic of two nodes. Recent efforts using HIN for recommendation represent each user, item and context entity as nodes with different types and can be categorized into two types. The first type leverages meta paths or knowledge-aware embeddings to encode the relatedness between users and items, and the relatedness are subsequently used to enhance the performance of recommendation\cite{yu2014personalized, shi2015semantic, zhao2017meta, hu2018leveraging, shi2018heterogeneous, shi2019semrec, zhang2016collaborative, ai2018learning, cao2019unifying}. The second type, as a comparison, due to the powerful representation ability of graph neural networks (GNNs), directly uses GNNs framework to obtain better user and item representations by aggregating neighborhood information\cite{berg2018graph, wang2019neural, fan2019metapath, wang2019kgat, wang2019kgcn, wang2019knowledge}.

\begin{figure}
\centering
\includegraphics[width=0.7\linewidth]{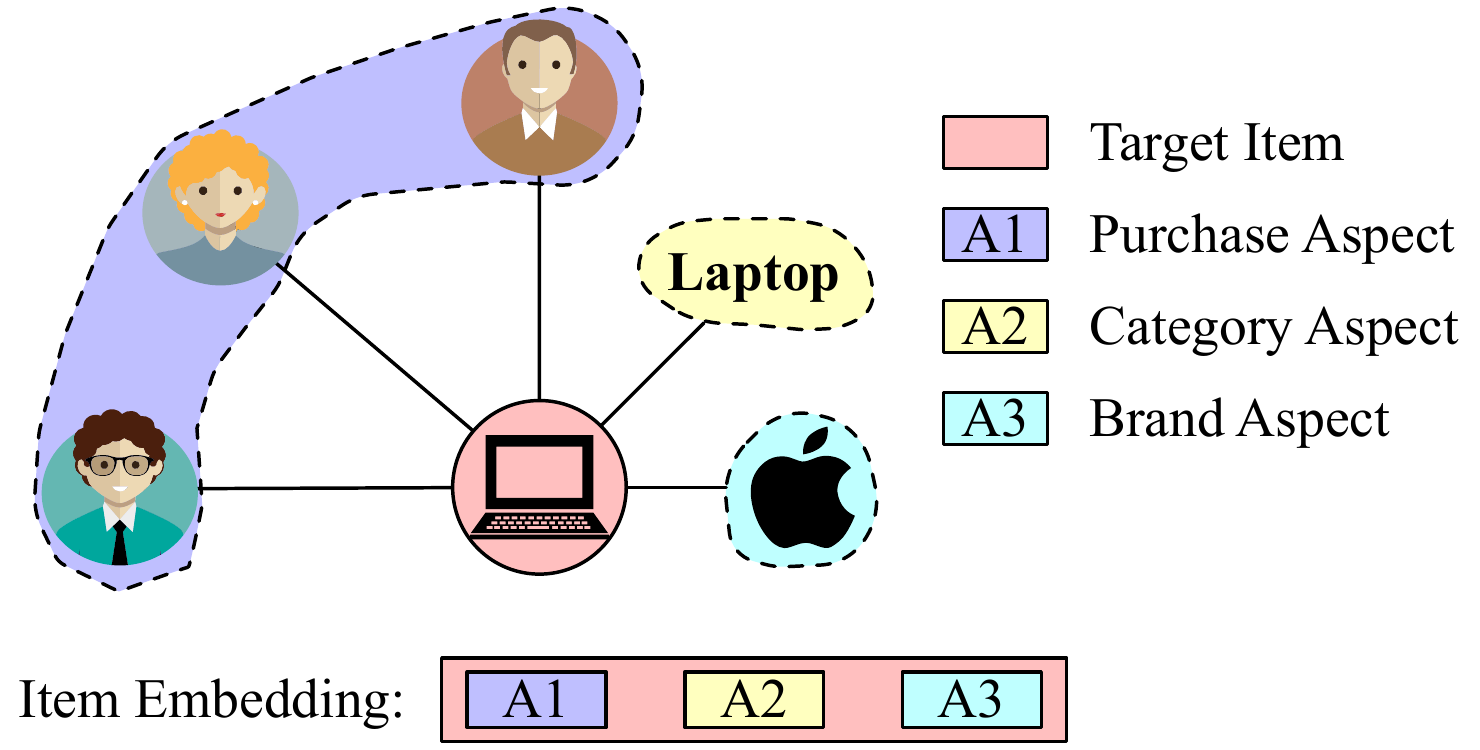}
\caption{An illustration of different aspects for the item representation.}
\label{fig:1}
\end{figure}
However, these studies share several common limitations. \textbf{First}, most of these works fail to disentangle the latent factors. As a toy example shown in Figure~\ref{fig:1}, the characteristics of Electronic products stem from several aspects. Since various types of connections reflect distinct aspects of a node, it is more natural and suitable to learn disentangled representations instead of preserving the confounding of the factors in HIN.
\textbf{Second}, previous works directly simplify meta paths as side information or connecting paths, neglecting the specific influences for each factor. In fact, different semantic relations have specific influence to the corresponding disentangled factors. For example, users have bought this product, reflecting the product is popular with certain groups; the product is laptop, which is suitable for office crowds; the brand is Apple and welcomed by the fans. 
\textbf{Third}, because of the above two points, these methods lead to non-robustness and low interpretability, i.e., they cannot explicitly model user preferences on the item with specific characteristics.
Thus, it is suitable to learn disentangled representations in HIN for recommendation. Particularly, user and item's representation can be disentangled into the same aspects. By calculating matching score of the target user and item in different aspects, user preferences on the item with specific characteristics can be explicitly modeled.



In this paper, we propose a novel \textbf{Disen}tangled \textbf{H}eterogeneous Graph \textbf{A}ttention \textbf{N}etwork (DisenHAN) for recommendation. Instead of exploiting entangled influential factors, our proposed approach explicitly models factors from different aspects and corresponding influence from semantic relations. Specifically, we represent user-item interactions with rich context information as a HIN, which consists different types of nodes and corresponding relations between node pairs. Given the node features as input, we project them into different embedding places and propose a semantic embedding propagation layer to identify the major aspect of the information flows in the relations, viewed as an iterative clustering process in the embedding space. Moreover, instead of designing meta paths manually, we take one-hop relations of nodes and use the attention mechanism to softly generate meta paths. By stacking multiple embedding propagation layers, each type of nodes especially users and items aggregate embeddings from different meta paths to capture the semantic information in high-order connectivity.

To summarize, in this paper we make the following contributions:
\begin{itemize}[leftmargin=*]
\item We propose to learn the disentangled representations with different aspects in HIN for recommendation. To the best of our knowledge, this is the first attempt to study the heterogeneous graph neural network on disentangled representations. And the proposed mechanism can be potentially generalized to other HIN based applications.
\item We propose a disentangled embedding propagation layer, which can iteratively identify the major aspect of the relation between node pairs and propagate corresponding information semantically. Moreover, we make the model more practical by automatically extract meta paths. 
\item We conduct extensive experiments on several real-world data sets to evaluate the proposed approach. Experimental results demonstrate that our approach not only outperforms the existing state-of-the-art models, but also has outstanding interpretability for recommendation.
\end{itemize}
\section{Related Work}
We review existing works on three lines of fields: 1) HIN based CF Methods, 2) Graph Neural Networks, 3) Disentangled Representation Learning.
\subsection{HIN based CF Methods}
HIN based CF methods have been proposed to avoid disparate treatment of each user-item interaction with rich context information. Especially meta paths in HIN, which are described as semantic relations between nodes pairs, can be defined by the network schema and used for recommendation. Shi et al. \cite{shi2015semantic, shi2019semrec} calculate the meta path based similarities between users, and infer the rating score based on similar users. Yu et al. \cite{yu2014personalized} and Zhao et al. \cite{zhao2017meta} employ various types of meta paths between users and items in matrix factorization to generate latent features for recommendation. Shi et al. \cite{shi2018heterogeneous} and Hu et al. \cite{hu2018leveraging} explicitly extract meta paths connecting users to items and integrate the representation of meta paths to improve recommendation. Han et al. \cite{han2018aspect} extract aspect-level factors from different meta paths and fuse the factors with attention mechanism for recommendation. However, defining effective meta paths require domain knowledge and is traditionally labor-intensive. Since knowledge graph (KG) can be seen as a special HIN, some works also leverage knowledge-aware embedding to guide the representation of items. Ai et al. \cite{ai2018learning} learn the knowledge graph embeddings to find connection paths between users and items. Cao et al. \cite{cao2019unifying} and Zhang et al. \cite{zhang2016collaborative} jointly learn latent representations in CF as well as items' semantic representations from KG. However, due to the lack of explicit encoding CF signals, which reveal behavior similarity of users/items, high order connectivity is hard to be captured\cite{wang2019neural}.   
\subsection{Graph Neural Networks}
Graph neural networks (GNNs)
, especially gated graph neural network (GGNN)\cite{li2016gated}, graph convolutional network (GCN)\cite{kipf2017semi}, graph inductive representation learning (GraphSAGE)\cite{hamilton2017inductive} and graph attention network (GAT)\cite{velivckovic2018graph}, have been attracting considerable attention recently. The key idea behind GNNs is aggregating features from nodes' neighbors via neural network. Compared to the traditional graph embedding models, the explicit and end to end manner exploiting high-order connectivity information in graph is suitable for the down-stream tasks, especially for recommendation. For example, GC-MC\cite{berg2018graph} and NGCF \cite{wang2019neural} leverage GNNs in user-item bipartite graph to CF. KGAT \cite{wang2019kgat}, KGCN \cite{wang2019kgcn} and KGCN-LS\cite{wang2019knowledge} extend items to knowledge graph, and utilize GNNs to strengthening items' representations. Although these works leverage GNNs for recommendation, they directly aggregate information from different types of nodes and ignore corresponding semantic relations. Recently, some HIN based GNNs are proposed for information propagation\cite{wang2019heterogeneous, fan2019metapath, zhang2019heterogeneous, hu2020heterogeneous}, but still overlook the different aspects of semantic information between nodes pairs. 
\subsection{Disentangled Representation Learning}
Disentangled representation learning, which aims to learn representations that separate explanatory factors of variations behind the data\cite{bengio2013representation}, has recently gained much attention. Not only such representations are demonstrated to be more robust, i.e., bring enhanced generalization ability as well as improved robustness to adversarial attack\cite{alemi2017deep}, but also make the downstream process more interpretable which can directly find applications in various fields with semantic data, such as images\cite{higgins2016beta, chen2016infogan, chen2017variational}, texts\cite{john2018disentangled} and user behaviors\cite{ma2019learning}. For graph-structure data, GAT performs multi-head attention to jointly attend information from different representation subspaces\cite{velivckovic2018graph}, while treats each subspace equally, and can be seen as a special case of disentangled representation learning. PolyDeepwalk \cite{liu2019single} notices multiple aspects of nodes, but learns the disentangled embeddings in separate two stages. Inspired by the capsule neural network (CapsNet)\cite{sabour2017dynamic}, some works replace scalar-valued neurons with vector-valued ones to learn disentangled node representations with GNNs' framework in homogeneous network\cite{verma2018graph, xinyi2018capsule, ma2019disentangled}. Our work focus on learning disentangled representation in HIN, a more complicated data structure with different node types and relations.

\section{Problem Definition}
In this section, we first introduce some important concepts related to HIN, meta path and meta relation. Next we give some formal definitions of our problem.

\textbf{Definition 3.1. Heterogeneous Information Network.}
A heterogeneous information network is defined as $\mathcal{G=(V,E)}$ with a node type mapping function $\mathcal{\phi:V\xrightarrow{}A}$ and an edge type mapping function $\mathcal{\psi:E\xrightarrow{}R}$, where each node $v \in\mathcal{V}$ and edge $e\in \mathcal{E}$ belong to one particular type in the corresponding node type set $\mathcal{A:\phi}(v)\in\mathcal{A}$ and edge type set $\mathcal{R:\psi}(e)\in\mathcal{R}$. Heterogeneous information networks have the property that $\mathcal{|A|+|R|}>2$.

\textbf{Definition 3.2. Meta Path and Meta Relation.}
A meta path is defined as a path in the form of $\mathcal{A}_1\xrightarrow{\mathcal{R}_1}\mathcal{A}_2\xrightarrow{\mathcal{R}_2}\dots\xrightarrow{\mathcal{R}_l}\mathcal{A}_{l+1}$ (abbreviated as $\mathcal{A}_1\mathcal{A}_2\dots \mathcal{A}_{l+1}$), which describes a composite relation $\mathcal{R}_1\circ\mathcal{R}_2\circ\dots\circ\mathcal{R}_l$ between node types $\mathcal{A}_1$ and $\mathcal{A}_{l+1}$, where $\circ$ denotes the composition operator on relations. However, a dataset-specific meta path always need to be selected manually based on prior knowledge from domain experts. With the number of meta paths growing largely, it becomes intractable to find optimal meta paths. 

Thus, we divide meta paths into individual meta relations and automatically generate meta paths based on different combination of meta relations. For an edge $e=(s,t)$ linked from source node $s$ to target node $t$, a meta relation is defined as $<\mathcal{\phi}(s),\mathcal{\psi}(e),\mathcal{\phi}(t)>$, which represents the semantic relation between the two nodes. Given target node $t$ as well as the meta relation $\mathcal{\psi}(e)$, there could be multiple source neighbors in type $\mathcal{\phi}(s)$. 

\textbf{Definition 3.3. Problem Definition.}
The HIN based recommendation problem can be formulated as follows. In a typical recommendation scenario, given a set of users $\mathcal{U}=\{u_1,u_2,...u_M\}$ and items $\mathcal{V}=\{v_1,v_2,...v_N\}$, we have user-item interaction matrix $Y\in Y^{M*N}$, where $y_{uv}=1$ indicates that user $u$ engages with item $v$, otherwise $y_{uv}=0$. Additionally, we also have corresponding context information of users and items, such as, social relationships between users, item brands and categories. We abstract user, item and their context information as different types of nodes connected by meta relations in $\mathcal{G}$. Our task is to predict whether user $u$ has a potential interest in item $v$ with which he has had no interaction before. Specially, we aim to learn the low-dimensional representations of $u$ and $v$ stem from different aspects for the prediction function $\hat{y}_{uv}=\mathcal{F}(u,v|\Theta,Y,\mathcal{G})$, where $\hat{y}_{uv}$ denotes the probability that user $u$ will engage with item $v$ and $\Theta$ denotes the model parameters of prediction function $\mathcal{F}$. 
\section{The Proposed Model}
In this section, we first introduce motivation and the overall architecture of the proposed model DisenHAN, which can iteratively identify the major aspect of meta relations for information propagation in HIN. 
We then present the details of each component and how they are applied to top-$N$ recommendation. 
\subsection{Overview}
\begin{figure*}
\centering
\includegraphics[width=0.8\textwidth]{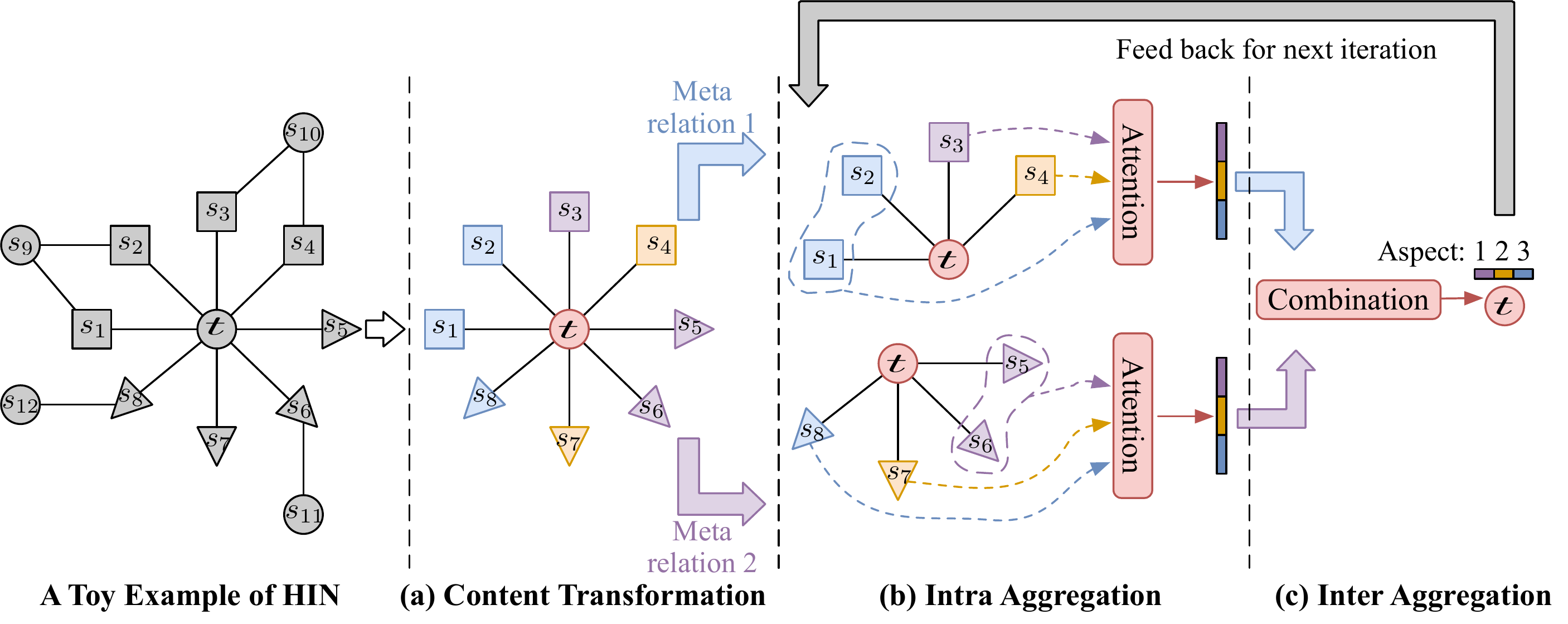}
\caption{A schematic view of the DisenHAN. Node shapes and colors indicate types and features, respectively. We assume that there are three aspects for the nodes. The same color between nodes and the relation denotes the major aspect of the relation where the grouped neighbors are similar in $k^{th}$ channel. The model iteratively takes features of target node $t$ as well as its neighbors as input and output three channels of disentangled representation for $t$. The output of the channels is feed back as features for the next iteration.} 
\label{fig:2}
\end{figure*}
The basic idea of our proposed model is to design a GNN to enrich the representation of users and items which are nodes in HIN. However, their representations often stem from different aspects and are determined by corresponding meta relations. Thus, previous GNNs are no longer suitable in this scenario.


As shown in Figure~\ref{fig:2}, there are three components in DisenHAN framework.
Given a target node $t$, we group its source neighbors based on meta relation $\mathcal{\psi}(e)$ and project their attributes into different subspaces to extract features from different aspects. Then, for each meta relation group, we aggregate source neighbor features under each aspect to capture the specific semantic information. Next, our proposed mechanism iteratively identifies the major aspect of each meta relation to get the optimal combined aspect information. We denote the output of $(l)$-th DisenHAN layer as $D^{(l)}$, which could be the input of the $(l+1)$-th layer. By stacking $L$ layers, we can aggregate multiple hops of neighbor information to enrich the representation of users and items, which are used for predicting the matching score. 
\subsection{Disentangled Content Transformation}
Different types of nodes always have attributes from different feature spaces in HIN. Thus, given target node $t$ and all its one hop neighbors $s\in N(t)$, we group them by different meta relations to get $N_{\mathcal{\psi}(e)}$, which denotes the same type of source nodes connecting to $t$ with meta relation $\mathcal{\psi}(e)$. Then for each group, we apply a type-specific transformation for both target and source nodes before feeding their features into DisenHAN layer. For $t$ of type $\mathcal{\phi}(t)\in\mathcal{A}$, we project its feature vector $x_t\in \mathbb{R}^{d_{in}}$ into $K$ different subspaces:
\begin{equation}
    c_{t,k} = \frac{\sigma(W_{\mathcal{\phi}(t),k}\cdot x_t)}{{\|\sigma(W_{\mathcal{\phi}(t),k}\cdot x_t)\|}_2},
\end{equation}
where $W_{\mathcal{\phi}(t),k}\in \mathbb{R}^{d_{in}\times\frac{d_{out}}{K}}$ is the weight of $k$-th channel, $\sigma$ is the activation function. Similarly, original features of $s$ are also projected into $K$ corresponding subspaces. Notice that for $t$, we can apply $L$ layers to aggregate information from multi-hops, and node features could be projected into different number of subspaces at different layers. In practice, local neighbors have more effect on the target node, thus, we add the constraint $k^{(1)}\geq K^{(2)}\geq\dots\geq K^{(L)}$ reflecting the number of aspects decrease along with neighbor hops increasing. 
\subsection{Disentangled Propagation Layer}
\subsubsection{Intra-relation Aggregation}
After type-specific transformation, the intra-relation layer learns the semantic information embedded in each meta relation $\mathcal{\psi}(e)$ by aggregating the features of grouped source nodes $N_{\mathcal{\psi}(e)}$. Despite the grouped neighbors are of the same type, different nodes still play an unequal role for each aspect. 

Inspired by the architecture design of Transformer\cite{vaswani2017attention}, we leverage attention mechanism to learn the weight of source nodes under different aspects. And the importance of neighbor $s\in N_{\mathcal{\psi}(e)}$ under aspect $k$ is calculated as:
\begin{equation}
\label{eqt:2}
    e_{t,s}^k = \text{ReLU}(\alpha_{\mathcal{\psi}(e)}^\mathrm{T}\cdot [z_{t,k}^{(i-1)}\parallel c_{s,k}]),
\end{equation}
where $\alpha_{\mathcal{\psi}(e)}\in \mathbb{R}^{2\frac{d_{out}}{K}}$ is attention vector for meta relation $\mathcal{\psi}(e)$. $z_{t,k}^{(i-1)}$, $c_{s,k}$ is the target and source node feature where we set $z_{t,k}^{(0)}=c_{t,k}$ initially, $\parallel$ denotes the concatenation operation. Considering different contribution of node importance under each aspect, we calculate the importance of the node by weighted summing all aspects:
\begin{equation}
\label{eqt:3}
    e_{t,s}^{\mathcal{\psi}(e)} = \sum_{k=1}^K r_{\mathcal{\psi}(e),k}^{(i-1)}\cdot e_{t,s}^k\ ,
\end{equation}
where $r_{\mathcal{\psi}(e),k}^{(i-1)}$ is the weight of aspect $k$ for meta relation $\mathcal{\psi}(e)$. We set $r_{\mathcal{\psi}(e),k}^{(0)}=1/K$ initially and the weight can be iteratively learned in inter-relation aggregation. After that, we can get the normalized attention weight for the grouped source nodes and aggregate their features under each aspect,
\begin{equation}
\label{eqt:4}
    \begin{split}
    \alpha_{t,s}^{\mathcal{\psi}(e)} &= \text{softmax}(e_{t,s}^{\mathcal{\psi}(e)}) = \frac{\exp(e_{t,s}^{\mathcal{\psi}(e)})}{\sum_{s'\in N_{\mathcal{\psi}(e)}}\exp(e_{t,s'}^{\mathcal{\psi}(e)})},\\
    z_{t,k}^{\mathcal{\psi}(e)} &= \sigma(\sum_{s\in N_{\mathcal{\psi}(e)}}\alpha_{t,s}^{\mathcal{\psi}(e)}\cdot c_{s,k}),
    \end{split}
\end{equation}
where $\sigma$ is the activation function. Notice that the attention weight is generated for single meta-relation, it is semantic-specific and able to capture corresponding semantic information. 
\subsubsection{Inter-relation Aggregation}
In this stage, we need to combine the semantic information revealed by all meta relations. Since each meta relation focuses on specific aspect, we should not fuse all the meta relations to describe the semantic information of aspect $k$. Therefore, we let the weight $r_{\mathcal{\psi}(e),k}$ be the probability that $\mathcal{\psi}(e)$ related to target node $t$ due to factor $k$,
which should satisfy $r_{\mathcal{\psi}(e),k}\geq0$ and $\sum_{k'=1}^K r_{\mathcal{\psi}(e),k} = 1$. For each aggregated aspect features, we learn the related weight as follows: 

\begin{equation}
\label{eqt:5}
    \hat{r}_{\mathcal{\psi}(e),k}^{(i)} = q_{\mathcal{\psi}(e)}^\mathrm{T}\cdot\tanh(W z_{t,k}^{\mathcal{\psi}(e)}),
\end{equation}
where $W$ is the weight matrix, $q_{\mathcal{\psi}(e)}$ is the semantic attention vector for meta relation $\mathcal{\psi}(e)$. And the related probability can be normalized by using softmax function,
\begin{equation}
\label{eqt:6}
    r_{\mathcal{\psi}(e),k}^{(i)} = \text{softmax}(\hat{r}_{\mathcal{\psi}(e),k}^{(i)}) = \frac{\exp(\hat{r}_{\mathcal{\psi}(e),k}^{(i)})}{\sum_{k'=1}^K\exp(\hat{r}_{\mathcal{\psi}(e),k'}^{(i)})}.
\end{equation}

Taking all the meta relations $\mathcal{\psi}_t$ connected to target node $t$, we iteratively search their major aspects and obtain the representation of $t$ as follows:
\begin{equation}
\label{eqt:7}
    z_{t,k}^{(i)} = \frac{c_{t,k} + \sum_{\mathcal{\psi}(e)\in \mathcal{\psi}_t}r_{\mathcal{\psi}(e),k}^{(i)}W z_{t,k}^{\mathcal{\psi}(e)}}{\|c_{t,k} + \sum_{\mathcal{\psi}(e)\in \mathcal{\psi}_t}r_{\mathcal{\psi}(e),k}^{(i)}W z_{t,k}^{\mathcal{\psi}(e)}\|_2},
\end{equation}
for iteration $i=1,2,\dots,I$, where the output $z_{t,k}^{(i)}$, $r_{\mathcal{\psi}(e),k}^{(i)}$ can in turn feed back to guide the attention weight for each meta relation grouped neighbors. The final enriched disentangled representation of target node $t$ can be denoted as $\{z_{t,1}^{(I)},z_{t,2}^{(I)},\dots,z_{t,K}^{(I)}\}$ and the overall iterative propagation process is shown in Algorithm 1. 
\begin{algorithm}
\caption{DisenHAN iterative forward propagation}
\KwIn {Features of target node $x_t\in \mathbb{R}^{d_{in}}$; all connected meta relations $\mathcal{\psi}_t$; features of meta relation grouped neighbors $\{x_s\in\mathbb{R}^{d_{in}}, \forall s\in N(t)\}$}
\KwOut{$\{z_{t,1},\dots,z_{t,K}\}$ (disentangled representation of $t$)}
\For{$v\in\{t\}\cup\{s:s\in N(t)\}$}{
\For{$k=1,2,\dots,K$}{
$c_{v,k}\leftarrow\sigma(W_{\mathcal{\phi}(v),k}\cdot x_v)$\;
$c_{v,k}\leftarrow c_{v,k}/\|c_{v,k}\|_2$\;
}
}
$z_{t,k}^{(0)}\leftarrow c_{t,k}, \forall k=1,2,\dots,K$\tcp*[r]{Initialize $K$ channels} 
\For{forward propagation iteration $i=1,2,\dots,I$}{
    \For{each meta relation $\mathcal{\psi}(e)\in\mathcal{\psi}_t$}{
        \For{$k=1,2,\dots,K$}{
            \For{$v\in N_{\mathcal{\psi}(e)}$}{Calculate $e_{t,s}^k$ using $z_{t,k}^{(i-1)}$\tcp*[r]{Eq.\ref{eqt:2}} Get attention weight using $r_{\mathcal{\psi}(e),k}^{(i-1)}$\tcp*[r]{Eq.\ref{eqt:3}}}
        Attention aggregate getting $z_{t,k}^{\mathcal{\psi}(e)}$ \tcp*[r]{Eq.\ref{eqt:4}}}
        Update getting $r_{\mathcal{\psi}(e),k}^{(i)}, \forall k=1,2,\dots,K$  \tcp*[r]{Eq.\ref{eqt:5}-\ref{eqt:6}}}
        \For{$k=1,2,\dots,K$}{
        $z_{t,k}^{(i)} = c_{t,k} + \sum_{\mathcal{\psi}(e)\in \mathcal{\psi}_t}r_{\mathcal{\psi}(e),k}^{(i)}W z_{t,k}^{\mathcal{\psi}(e)}$\;
        $z_{t,k}^{(i)} = z_{t,k}^{(i)}/\|z_{t,k}^{(i)}\|_2$ \tcp*[r]{Update in Eq.\ref{eqt:7}}
    }
}
$z_{t,k}\leftarrow z_{t,k}^{(I)}, \forall k=1,2,\dots,K$
\end{algorithm}
\vspace{-10pt}
\subsection{Recommendation}
We treat the user as target node and the output of $L$ layers of disentangled propagation is a set of embedding vectors $\{z_{u,1},z_{u,2}\dots,z_{u,K}\}$, representing different aspects of user. Similarly, we can treat the item as target node and get the corresponding $K$ aspects of item embedding vectors denoted as $\{z_{v,1},z_{v,2}\dots,z_{v,K}\}$. For the final top-$N$ recommendation, we simply conduct the inner product of each aspect to estimate the user's preference towards the target item in aspect $k$, and sum all the aspects as the final matching score:
\begin{equation}
    s_{uv} = \sum_{k=1}^K z_{u,k}^\mathrm{T}\cdot z_{v,k}.
\end{equation}
\subsection{Training}
To optimize the top-$N$ recommendation with implicit feedback, we optimize it as a binary classification problem. The prediction score $\hat{y}_{uv}$ then represents how likely user $u$ will select item $v$ in the future. We constrain the matching score $s_{uv}$ in the range of $[0,1]$, which can be estimated by using a probabilistic function like Logistic,
\begin{equation}
    \hat{y}_{uv} = \text{sigmod}(s_{uv}) = \frac{1}{1+e^{-s_{uv}}}.
\end{equation}
Over all the training set, we learn the parameters of the model with negative sampling and the complete loss function is as follows:
\begin{equation}
    \mathcal{L} = -\sum_{(u,v)\in\mathcal{Y}}\log\hat{y}_{uv} - \sum_{(u,v^-)\in\mathcal{Y^-}}\log(1-\hat{y}_{uv^-}),
\end{equation}
where $\mathcal{Y}$ is positive instances and $\mathcal{Y}^-$ is negative instances uniformly sampled from unobserved interactions by controlling the sampling ratio w.r.t. number of observed interactions.

We leverage mini-batch training to calculate the gradient. For each training iteration, we consider $B$ size of user-item pairs. And for each mini-batch, we need to fix the number of source nodes to the max and pad zeros due to the different number of nodes in each meta relation. However, we observe that the number of nodes in some relations tends to obey a long-tail distribution, so the padding would lead to both high space and time complexity. Followed by the previous work\cite{hamilton2017inductive}, we sample fixed number of source nodes and pad zeros when source nodes less than the fixed number.

\subsection{Analysis of DisenHAN}
Given target node and meta relation grouped source neighbors, we now analyze the iterative propagation process in HIN.

\textbf{Iterative Forward Propagation Process.}
The key insight here is that we should only combine the feature from related meta relations to enrich the aspect $k$ of target node. Thus, the task of the process is to design a mechanism inferring the related source neighbors under aspect $k$. We suppose aspect $k$ is likely to be the major aspect of the meta relation if grouped source neighbors are similar w.r.t. aspect $k$, i.e., they form a cluster in the $k^{th}$ subspace. Thus, the iterative forward propagation is actually a clustering process which searches the major cluster of the grouped neighbors and aggregates their features to the target node. We calculate the aspect weight of each meta relation for each iteration, which can attend source nodes in the largest cluster to prune noise and achieve fast convergence. 

\textbf{Convergence Analysis.} 
The iterative forward propagation process is equivalent to an expectation maximization(EM) algorithm for the mixture model. Let $Z=\{z_{t,k}\}_{k=1}^K$ as estimated representation of target $t$, $C=\{c_{v,k}:v\in \{t\}\cup\{s:s\in N(t)\}\}$ as noisy observation of $Z$ and $R=r_{\mathcal{\psi}(e),k}$ for each meta relation. The EM algorithm maximize $p(C;Z)=\sum_R p(C,R;Z)$, and the log-likelihood can be formulated as
\begin{equation}
    \ln P(C;Z) = \sum_R q(R)\ln \frac{P(C,R;Z)}{q(R)} + \sum_R q(R)\ln\frac{q(R)}{p(R|C;Z)},
\end{equation}
where the first term is evidence lower bound denoted by $\text{ELBO}(C;R,Z)$ and second term is the Kullback-Leibler (KL) divergence with non-negative value. Our model alternatively updates $q(R)$ and $Z$ by a) setting optimal $q(R)$ to $p(R|C;Z)$ so that $\text{ELBO}(C;R,Z) = \ln P(C;Z)$ for current $C$ and $Z$ in E step and b) maximizing ELBO w.r.t. $Z$ in M step. Thus, for $i^{th}$ iteration, $\ln P(C;Z^{(i)}) \geq \text{ELBO}(C;R^{(i)},Z^{(i)}) \geq \text{ELBO}(C;R^{(i)},Z^{(i-1)}) = \ln P(C;Z^{(i-1)})$, which improves the log-likelihood monotonically and the algorithm converges. 

\textbf{Automatic Meta Paths Selection.} 
Our model can iteratively identify aspect weight of each meta relation which can be seen as the semantic relatedness among nodes. For each layer, we incorporate one hop neighbor information. And by stacking multiple layers, the proposed attention mechanism is able to automatically aggregate corresponding aspect information multiple hops away, which can be seen as a soft way to select meta paths.  


\section{Experiments}
In this section, we present experiments to evaluate the effectiveness of our proposed approach. We aim to answer the following research questions:

\textbf{RQ1:}
How does DisenHAN perform compared with state-of-the-art models for top-$N$ recommendation? Can DisenHAN has the potential to alleviate cold-start problems?


\textbf{RQ2:}
Is it necessary to keep key components of DisnHAN, such as disentangled representation, meta relation aspect weight iterative calculation? How do hyper-parameters in DisenHAN impact recommendation performance?

\textbf{RQ3:}
Will DisenHAN interpret different semantics of meta relations in HIN for recommendation?

We first present the experimental settings, followed by answering the above three research questions.

\subsection{Experimental Settings}
\subsubsection{Data Sets}
We apply our model to three public accessible data sets for top-$N$ recommendation. The statistics of the data sets are summarized in Table~\ref{tab:1}.

\textbf{Yelp}\footnote{{\url{https://www.yelp.com/dataset/}}}
This is a widely adopted local business recommendation data set, where users can review local businesses online. We treat local businesses like restaurants as the items and each review as an interaction between the user and item. To ensure the quality of the data set, we use the core setting, i.e., retaining users and items with at least ten interactions and users with at least five friends. 

\textbf{Amazon-electricity}\footnote{{\url{http://jmcauley.ucsd.edu/data/amazon/}}}
Amazon-review is a widely used data set for product recommendation\cite{he2016ups}. We select Amazon-electricity from the collection and treat each review as an interaction between the user and item. Similarly, we use the core setting to ensure that each user and item have at least eight interactions and each item has at least two related items. 

\textbf{MovieLens}\footnote{{\url{https://grouplens.org/datasets/hetrec-2011/}}}
This is a widely used benchmark data set in movie recommendation, which consists of explicit ratings on the MovieLens website. We follow the prior work\cite{he2017neural} to transform it into implicit data, where each interaction between the user and item is marked as 0 or 1 indicating whether the user has rated the item or not. We treat movies as the items and use the core setting where each user and item have at least five interactions in order to ensure data quality.

For each data set, we rank the interactions in chronological order and select the first 80\% of historical interactions as the training set with the remaining 10\%, 10\% as the validation and test data set respectively.
\begin{table}
\centering
\caption{Descriptive statistics of our three datasets.}
\label{tab:1}
\label{tab:statics}
\setlength{\tabcolsep}{3pt}
\begin{tabular}{l||l|l|l} 
\toprule 
\textbf{Dataset} & Node & Relation & Avg.Degree \\
\midrule 
\multirow{4}{*}{Yelp} 
& \#user(U): 32,654 & \#U-B: 1,347,861 & \#U-B: 41.3/39.4\\
& \#business(B): 34,193 & \#U-U: 1,566,400 & \#U-U: 48.0/48.0\\
& \#city(Ci): 387 & \#B-Ci:34,193 & \#B-Ci: 1.0/88.4\\
& \#category(Ca): 1006 & \#B-Ca: 156,886 & \#B-Ca: 4.6/156.0\\
\midrule
\multirow{4}{*}{Amazon} 
& \#user(U): 13,982 & \#U-I: 181,118 & \#U-I: 13,0/25.7\\
& \#item(B): 7,044 & \#I-I: 246,050 & \#I-I: 34.9/34.9\\
& \#brand(Br): 1,151 & \#I-Br: 7,044 & \#I-Br: 1.0/6.1\\
& \#category(Ca): 519 & \#I-Ca: 30,804 & \#I-Ca: 4.4/59.4\\
\midrule
\multirow{6}{*}{Movielens} 
& \#user(U): 2,113 & \multirow{6}{*}{\tabincell{l}{\#U-M: 855,598 \\ \#M-A: 95,777 \\ \#M-D: 10,068 \\ \#M-C: 10,109 \\ \#M-G: 20,670 }} & \multirow{6}{*}{\tabincell{l}{\#U-M: 405.0/84.6 \\ \#M-A: 9.5/2.5 \\ \#M-D: 1.0/2.5 \\ \#M-C: 1.0/140.0 \\ \#M-G:  2.0/1033.5 }}\\
& \#movie(M): 10,109 &  & \\
& \#actor(A): 38,044 &  & \\
& \#director(D): 4,031 & & \\
& \#country(C): 72 & & \\
& \#genre(G): 20 & & \\
\bottomrule 
\end{tabular}
\end{table}

\begin{table*}
\centering
\caption{Results of effectiveness experiments on three different datasets.}
\label{tab:2}
\begin{tabular}{cccccccccc}
\toprule 
\multirow{2}{*}{Model} & \multicolumn{3}{c}{Yelp} & \multicolumn{3}{c}{Amazon} & \multicolumn{3}{c}{Movielens}\\ 
\cmidrule[0.5pt](lr){2-4}\cmidrule[0.5pt](lr){5-7}\cmidrule[0.5pt](lr){8-10}
& Prec@10 & Recall@10 & NDCG@10 & Prec@10 & Recall@10 & NDCG@10 & Prec@10 & Recall@10 & NDCG@10\\
\midrule 
BPR-MF & 0.2793 & 0.4734 & 0.5996 & 0.0780 & 0.2897 & 0.3126 & 0.4864 & 0.1518 & 0.6964\\
NeuMF & 0.2717 & 0.4463 & 0.5899 & 0.0811 & 0.2791 & 0.3164 & 0.6016 & 0.1710 & 0.7838\\ 
\midrule
$\text{FMG}_{\text{rank}}$ &0.2810 & 0.4749 & 0.6055 & 0.0788 & 0.2717 & 0.3120 & 0.5642 & 0.1460 & 0.7375 \\
MCRec & 0.2874 & 0.4642 & 0.6023 & 0.0753 & 0.2882 & 0.3048 & 0.5477 & 0.1477 & 0.7511\\
NeuACF & 0.2928 & 0.4861 & 0.6215 & 0.0714 & 0.2686 & 0.2833 & 0.5855 & 0.1682 & 0.7719\\
\midrule
GC-MC & 0.2819 & 0.4750 & 0.5983 & 0.0799 & 0.2760 & 0.3026 & 0.4964 & 0.1508 & 0.7061\\
NGCF & 0.2853 & 0.4803 & 0.6075 & 0.0754 & 0.2829 & 0.2953 & 0.5533 & 0.1662 & 0.7532\\
HAN & 0.3102 & 0.4937 & 0.6227 & 0.0830 & 0.2872 & 0.3166 & 0.5770 & 0.1524 & 0.7681\\
\midrule
DisenHAN(Ours) & \textbf{0.3174} & \textbf{0.5117} & \textbf{0.6512} & \textbf{0.0859} & \textbf{0.3193} & \textbf{0.3451}& \textbf{0.6145} & \textbf{0.1761} & \textbf{0.8000}\\
\bottomrule 
\end{tabular}
\end{table*}
\subsubsection{Evaluation Metrics}
For each user in the test data set, we treat all the items that the user has not interacted with as the negative items. Since it is time consuming to rank all items for every user during evaluation, we follow the common strategy that randomly samples 100 negative items and rank the test items among the 100 items for each user\cite{he2017neural, hu2018leveraging}. To evaluate the performance of top-$N$ recommendation for all methods, we adopt three widely used metrics: Precision, Recall and Normalized Discounted Cumulative Gain(NDCG). By default, we truncate the ranked list at 10 for both metrics, which are Prec@10, Recall@10 and NDCG@10. 

\subsubsection{Baselines}
To demonstrate the effectiveness, we compare DisenHAN with three classes of methods: (A) classical CF methods; (B) HIN based recommenders, which model user-item interaction with rich context information as HIN; and (C) GNN based recommendation methods, which use information propagation way to exploit high-order connectivity in graph.  
\begin{itemize}[leftmargin=*]
\item \textbf{BPR-MF}~\cite{rendle2009bpr} (A): This is a matrix factorization model optimized by the Bayesian personalized ranking (BPR), which exploits user-item implicit feedback directly.
\item \textbf{NeuMF}~\cite{he2017neural} (A): This method combines the linearity of matrix factorization and non-linearity of deep neural networks for modeling the user-item interactions.
\item \textbf{FMG}$_{rank}$~\cite{zhao2017meta} (B): This is a HIN based method which combines meta graph based matrix factorization with factorization machine for rating prediction. we follow the prior work\cite{hu2018leveraging} to modify its optimization objective as BPR for top-$N$ recommendation.
\item \textbf{MCRec}~\cite{hu2018leveraging} (B): This method leverages meta-path based context with co-attention mechanism for top-$N$ recommendation in HIN.
\item \textbf{NeuACF}~\cite{han2018aspect} (B): This method considers multiple aspects of users and items with a deep neural network for recommendation in HIN.
\item \textbf{GC-MC}~\cite{berg2018graph} (C): This model adopts GCN\cite{kipf2017semi} encoders in user-item bipartite graph to generate the representation of users and items, where only the first-order neighbors are considered.
\item \textbf{NGCF}~\cite{wang2019neural} (C): This model uses embedding propagation layer to enrich user and item representations for exploiting high order connectivity in user-item bipartite graph. 
\item \textbf{HAN}~\cite{wang2019heterogeneous} (C): This is a heterogeneous graph neural network with attention mechanism to aggregate neighbor information via different meta paths. We modify the model for top-$N$ recommendation.
\end{itemize}

\subsubsection{Implementation Detail}
We implement our DisenHAN model in Pytorch. The embedding size is fixed to 100 for all models. For our method, we set two propagation layers and the number of iterations $I$ per layer is 5 in default setting. Within the first and second layers, we set the number of aspects as 5, corresponding embedding dimension is 20 per aspect. To prevent overfitting, we employ dropout where the ratio is tuned amongst $\{0.0,0.1,\dots,0.9\}$.  We optimize DisenHAN with Adam optimizer by setting the learning rate to 0.005 and using early stopping with a patience of 20, i.e. we stop training if recall@10 on the validation set does not increase for 20 successive epochs. For baseline methods, we split exactly the same training, validation and test set as DisenHAN and apply gird search for optimal hyper-parameters.


\subsection{Performance Comparison(RQ1)}
\subsubsection{Overall Comparison}
We summarize the results by comparing the performance of all the methods shown in table~\ref{tab:2} and have the following observations.
\begin{itemize}[leftmargin=*]
\item HIN based recommendation methods, FMG$_{rank}$, MCRec and NeuACF, utilize designed meta paths to capture semantic relations between user and item node pairs. By modeling this context information, the performance of HIN based recommenders improves compared to the classical CF methods in most cases. However, due to the lack of explicit encoding CF signals, high-order connectivity of users and items are still hard to be captured and the improvement is marginal. Moreover, meta paths selection in these HIN based methods is more rely on prior knowledge from domain experts which can have a great effect on the performance.
\item GNN based methods achieve better performance than classical CF methods in most cases, which indicates the positive effect of modeling high-order connectivity based on information propagation. However, GC-MC only utilizes the first-order neighbors to guide the information aggregation, and NGCF utilizes multiple-order neighbors without semantic relations in HIN. HAN models information propagation on meta path based neighbors. Compared with the GC-MC and NGCF, HAN improves the recommendation performance which verifies the importance of capturing collaborative signals by modeling semantic relations in HIN. 
\item DisenHAN performs consistently better than other baselines on all the datasets. In particular, DisenHAN improves over the strongest baselines w.r.t. Prec@10 by  $2.3\%$, $3.5\%$, $2.1\%$; Recall@10 by $3.6\%$, $10.8\%$, $3.0\%$ and NDCG@10 by $4.6\%$, $9.0\%$, $2.1\%$ in Yelp, Amazon-electricity and Movielens respectively. By stacking multiple disentangled propagation layers, DisenHAN is capable of exploring high-order connectivity in HIN with an explicit way, so as to capture collaborative signal effectively. Compared with HIN based and GNN based methods, DisenHAN can capture different aspects of semantic information and the significant improvement indicates the positive effect to achieve better representations for recommendation. 
\end{itemize}

\subsubsection{Cold-start Recommendation}
\begin{figure*}
\centering
\begin{subfigure}[t]{0.32\linewidth}
\includegraphics[width=\linewidth]{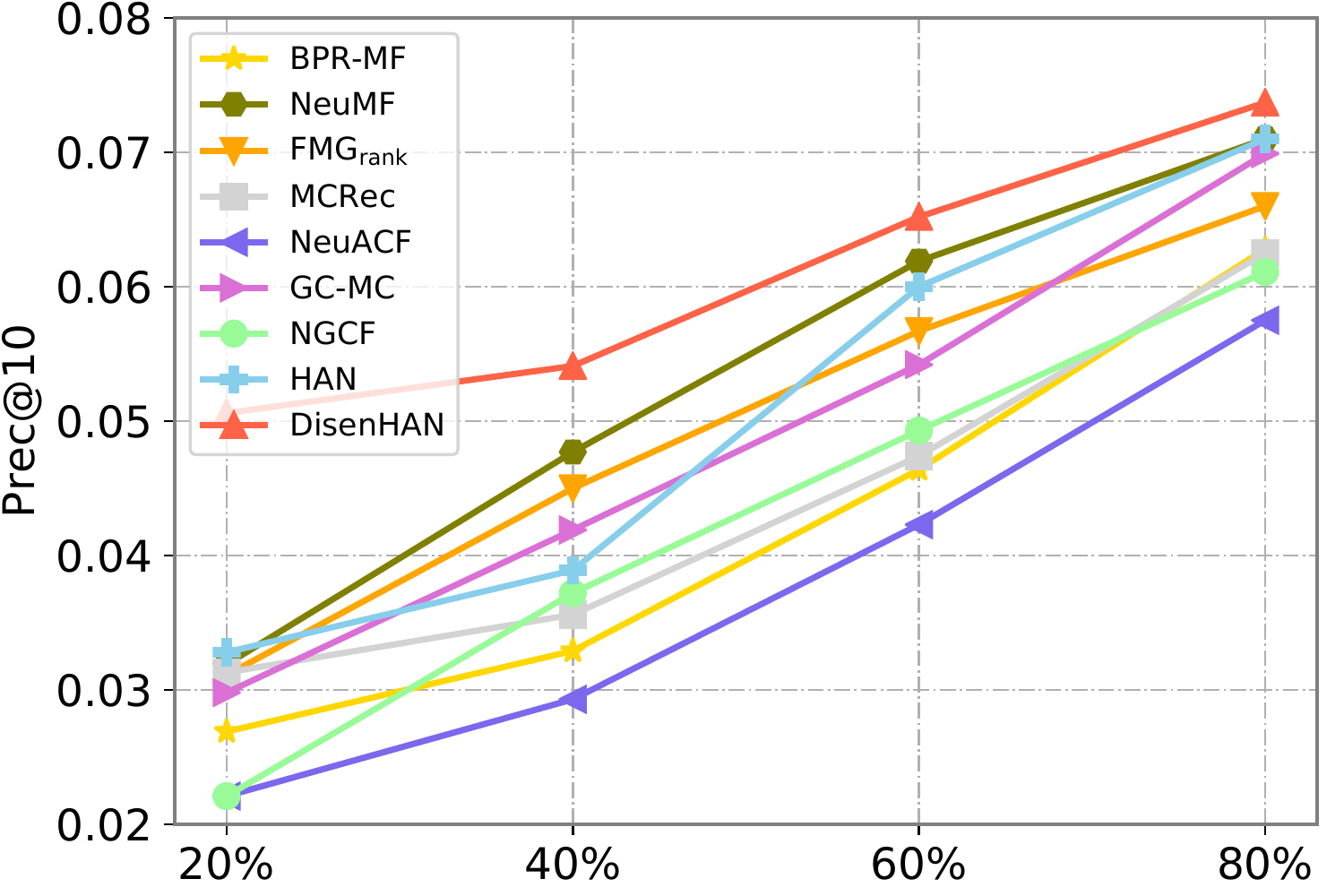}
\caption{Prec@10} \label{fig:3_a}
\end{subfigure}
\hspace*{\fill} 
\begin{subfigure}[t]{0.32\linewidth}
\includegraphics[width=\linewidth]{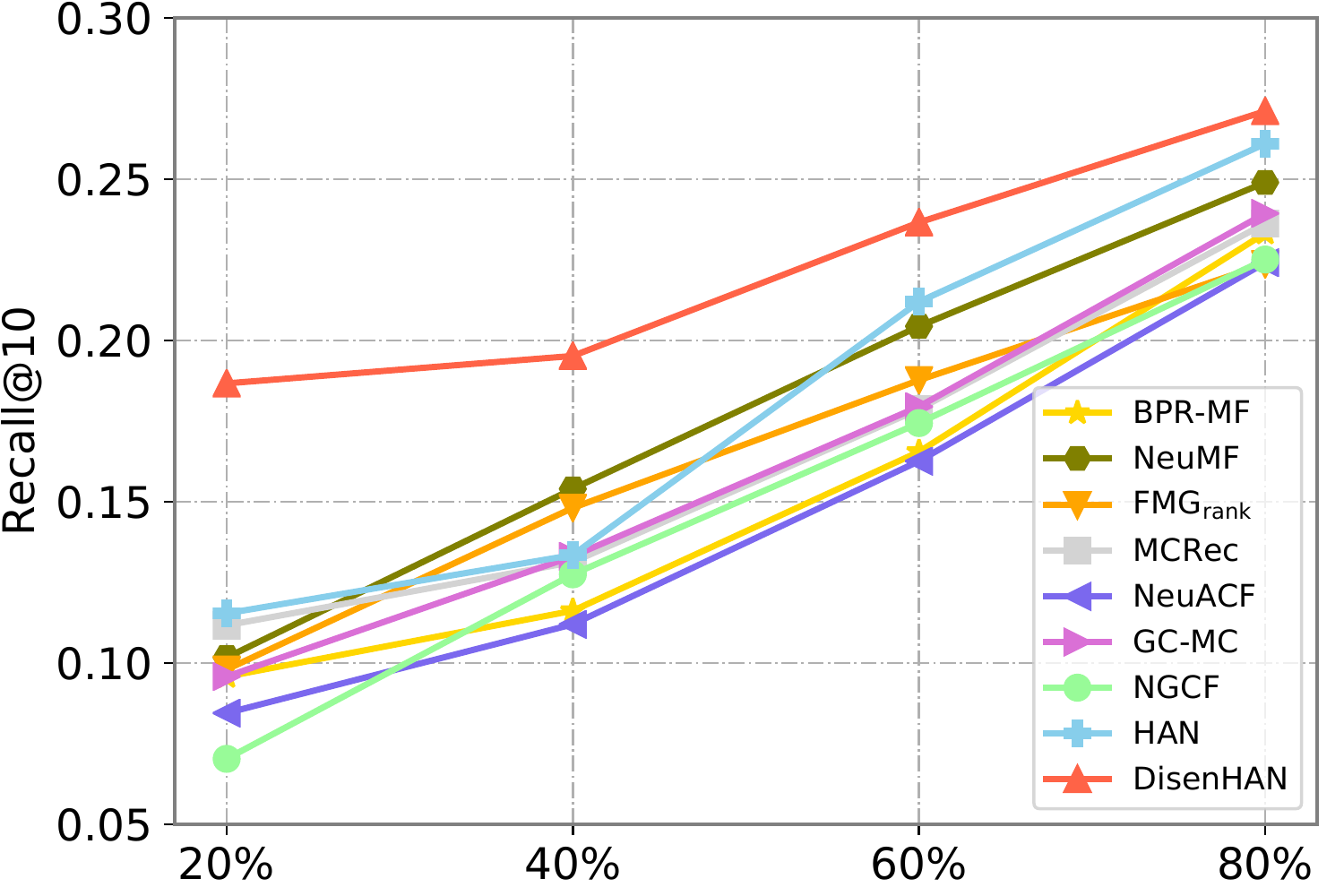}
\caption{Recall@10} \label{fig:3_b}
\end{subfigure}
\hspace*{\fill} 
\begin{subfigure}[t]{0.32\linewidth}
\includegraphics[width=\linewidth]{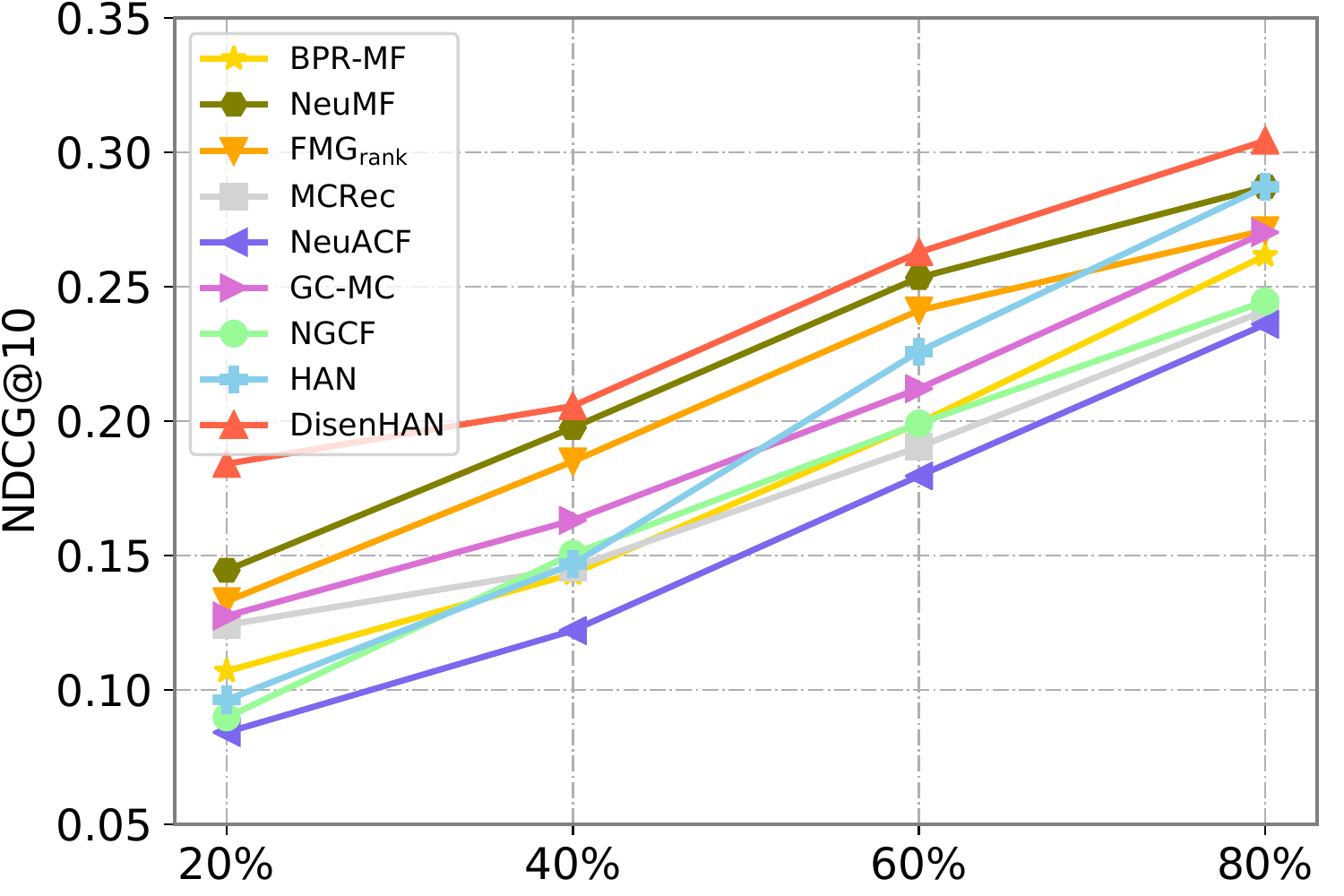}
\caption{NDCG@10} \label{fig:3_c}
\end{subfigure}
\caption{Performance comparison over the sparsity distribution of data on \textit{Amazon}.}
\label{fig:3}
\end{figure*}

The sparsity issue usually limits the performance of recommender systems since few user-item interactions are insufficient to generate high-quality representations. HINs are particularly useful to alleviate the problem by considering the context information of users and items. We study whether DisenHAN can alleviate this issue. Towards this end, we divide the training data into five equal folds and vary the amount of training data from one fold to four folds, corresponding to $20\%$, $40\%$, $60\%$, $80\%$ of entire training data as training sets. Figure~\ref{fig:3} illustrates the performance w.r.t. different sparsity distribution of data in Amazon data set. Yelp and Movielens data sets are similar and omitted due to the space limitation. We find that:
\begin{itemize}[leftmargin=*]
    \item DisenHAN consistently outperforms other baselines over the sparsity distribution of data, and the improvement over other baselines decreases with the increasing of training data. It indicates that our model can effectively improve the recommendation performance especially when the interaction records are sparse.
    \item GNN based recommendation methods, i.e., NGCF, though captures the high connectivity information between users and items, can not guarantee the improvement over classical recommendation methods. This may attribute to the fact that they ignore the different aspects of semantic relations between users and items.
\end{itemize}

\subsection{Study of DisenHAN(RQ2)}
\begin{figure}
\centering
\begin{subfigure}[t]{0.49\linewidth}
\includegraphics[width=\linewidth]{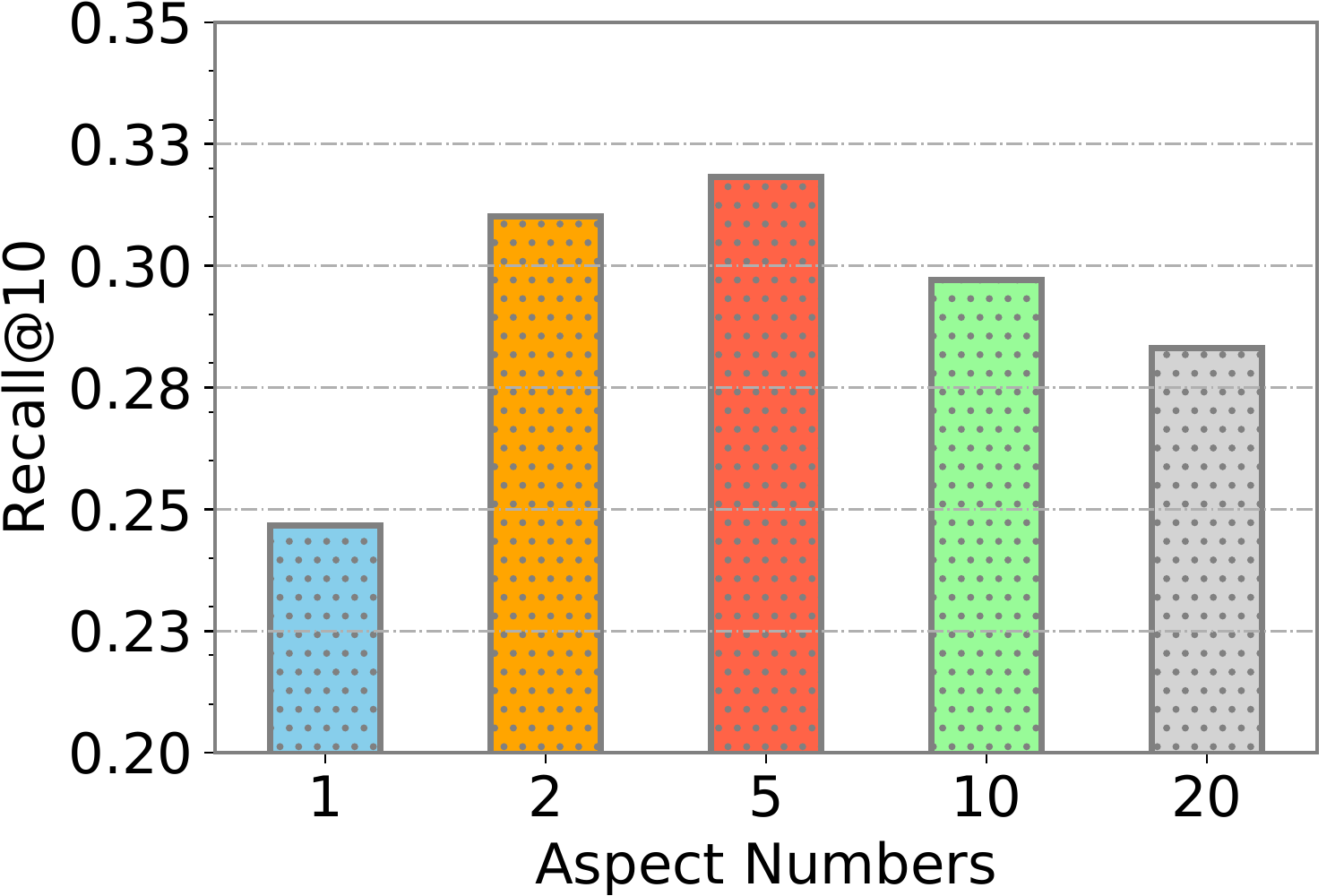}
\caption{Recall@10} \label{fig:4_b}
\end{subfigure}
\hspace*{\fill} 
\begin{subfigure}[t]{0.49\linewidth}
\includegraphics[width=\linewidth]{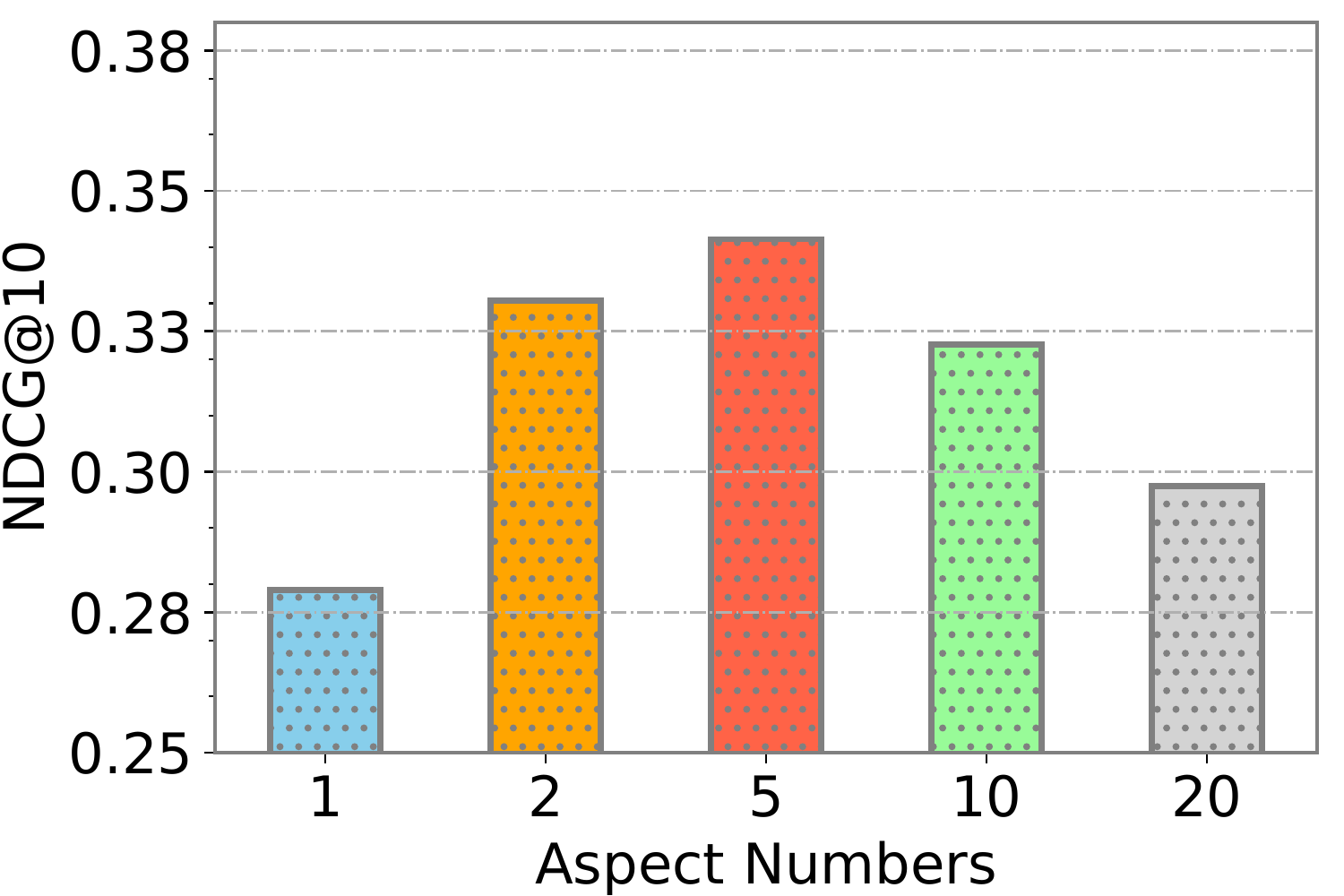}
\caption{NDCG@10} \label{fig:4_c}
\end{subfigure}
\caption{Performance w.r.t. different aspect numbers on \textit{Amazon}.}
\label{fig:4}
\end{figure}

As the disentangled propagation layer plays an important role in DisenHAN, we investigate its impact on the performance. We start by exploring the effect of aspect numbers for disentangled propagation layer, then we study how iteration times of each layer affect the performance. At last, we stack different numbers of disentangled propagation layers and compare the variations of our model.
\subsubsection{Effect of Disentangled Representation}
To investigate whether or not DisenHAN can benefit from disentangled representation, we study the performance of the model with varying aspect numbers in one disentangled propagation layer. In particular, we search the aspect numbers in the range of $\{1,2,5,10\}$. Figure~\ref{fig:4} summarizes the experimental results w.r.t. the Amazon data set and we have the following observations:
\begin{itemize}[leftmargin=*] 
    \item When the number of aspects for the representation is 1, the model can be degraded into HIN based GNN model without disentangled representation. The performance of the model is poor, which indicates modeling different aspects of semantic information can greatly facilitate the recommendation task.
    \item Increasing the number of aspects for disentangled propagation layer can substantially enhance the recommendation performance. And DisenHAN performs best when the number of aspects is around 5, which is close to the actual number in Amazon data set.
    \item However, when the number of aspects is very large, i.e., $K>10$, the improvement is much lower which might be caused by applying too complex semantic structure for the graph.
\end{itemize}

\subsubsection{Effect of Iteration Times for Layer}
\begin{figure}
\centering
\begin{subfigure}[t]{0.49\linewidth}
\includegraphics[width=\linewidth]{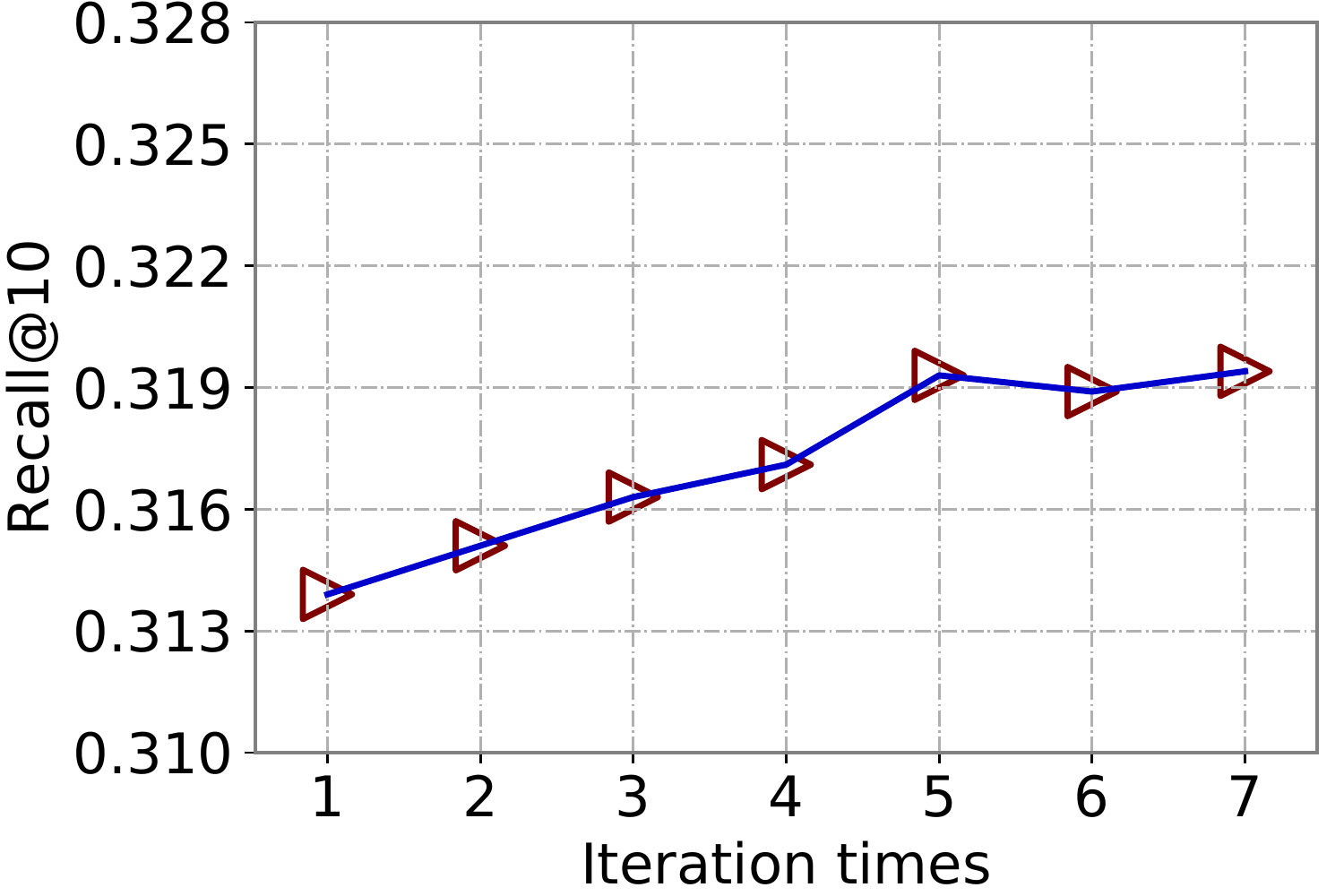}
\caption{Recall@10} \label{fig:5_b}
\end{subfigure}
\hspace*{\fill}
\begin{subfigure}[t]{0.49\linewidth}
\includegraphics[width=\linewidth]{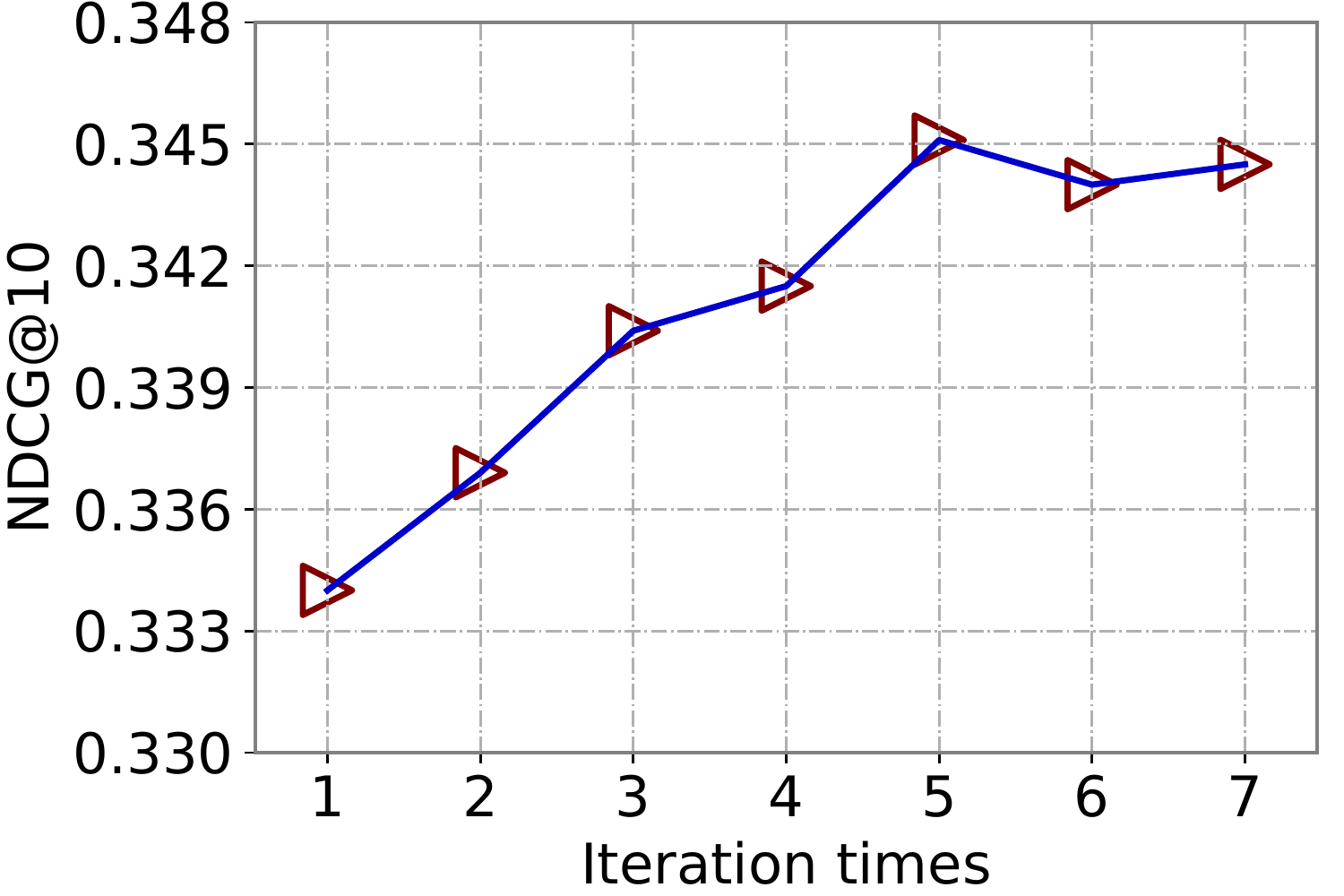}
\caption{NDCG@10} \label{fig:5_c}
\end{subfigure}
\caption{Performance w.r.t. different iteration times on \textit{Amazon}.}
\label{fig:5}
\end{figure}
To further investigate how the iteration times affect the performance, we fix the aspect number of one disentangled propagation layer at 5 and consider the variants of the model with different iteration times. As shown in Figure~\ref{fig:5}, we set the iteration times in the range of $[1,7]$ and we have the following observations:
\begin{itemize}[leftmargin=*]
    \item More iterations generally leads to better performance before saturation which is guaranteed by its convergence properties. When the iteration times reach to 5, the performance becomes stable showing that few iteration times is enough to get the satisfactory performance.    
    \item Compared with other hyper parameters, i.e., the number of aspects, DisenHAN is not sensitive to the iteration times. This may be that multiple epochs of training can also make the model with smaller iteration times converge.  
\end{itemize}

\subsubsection{Effect of Propagation Layer Number}
\begin{table}
\centering
\caption{Effect of disentangled propagation layer numbers}
\label{tab:3}
\begin{tabular}{ccccc}
\toprule 
Data sets & Layers & Prec@10 & Recall@10 & NDCG@10\\
\midrule
\multirow{3}{*}{Yelp} & 1 & 0.3120 & 0.5051 & 0.6528\\
& 2 & 0.3174 & 0.5117 & 0.6512\\
& 3 & 0.3136 & 0.5059 & 0.6444\\
\midrule 
\multirow{3}{*}{Amazon} & 1 & 0.0857 & 0.3182 & 0.3414\\
& 2 & 0.0859 & 0.3193 & 0.3451\\
& 3 & 0.0858 & 0.3185 & 0.3398\\
\midrule 
\multirow{3}{*}{Movielens} & 1 & 0.6084 & 0.1748 & 0.7869\\
& 2 & 0.6145 & 0.1761 & 0.8000\\
& 3 & 0.6112 & 0.1741 & 0.7932\\
\bottomrule 
\end{tabular}
\end{table}

By stacking different numbers of propagation layers, we investigate how the depth of DisenHAN affects the performance. In particular, we stack the layer numbers in the range of $\{1,2,3\}$. Table~\ref{tab:3} summarizes the experimental results and we have the following observations:
\begin{itemize}[leftmargin=*]
    \item More disentangled propagation layers will yield influence from high-order neighbors of the target node and increase the depth of model. Clearly, 2 layers of model achieve better performance than 1 layer. It means by stacking multiple propagation layers, we can automatically select the corresponding meta paths for each data set and aggregate information from meta paths based neighbors.
    \item When stacking more than 2 propagation layers, the influence of the neighbors on the performance is small and can lead to overfitting. This is reasonable because short meta-paths is good enough \cite{sun2011pathsim} and deep architecture might introduce noises to the representation learning.
\end{itemize}

\subsection{Case Study and Visualization(RQ3)}

\begin{figure}
\centering
\includegraphics[width=\linewidth]{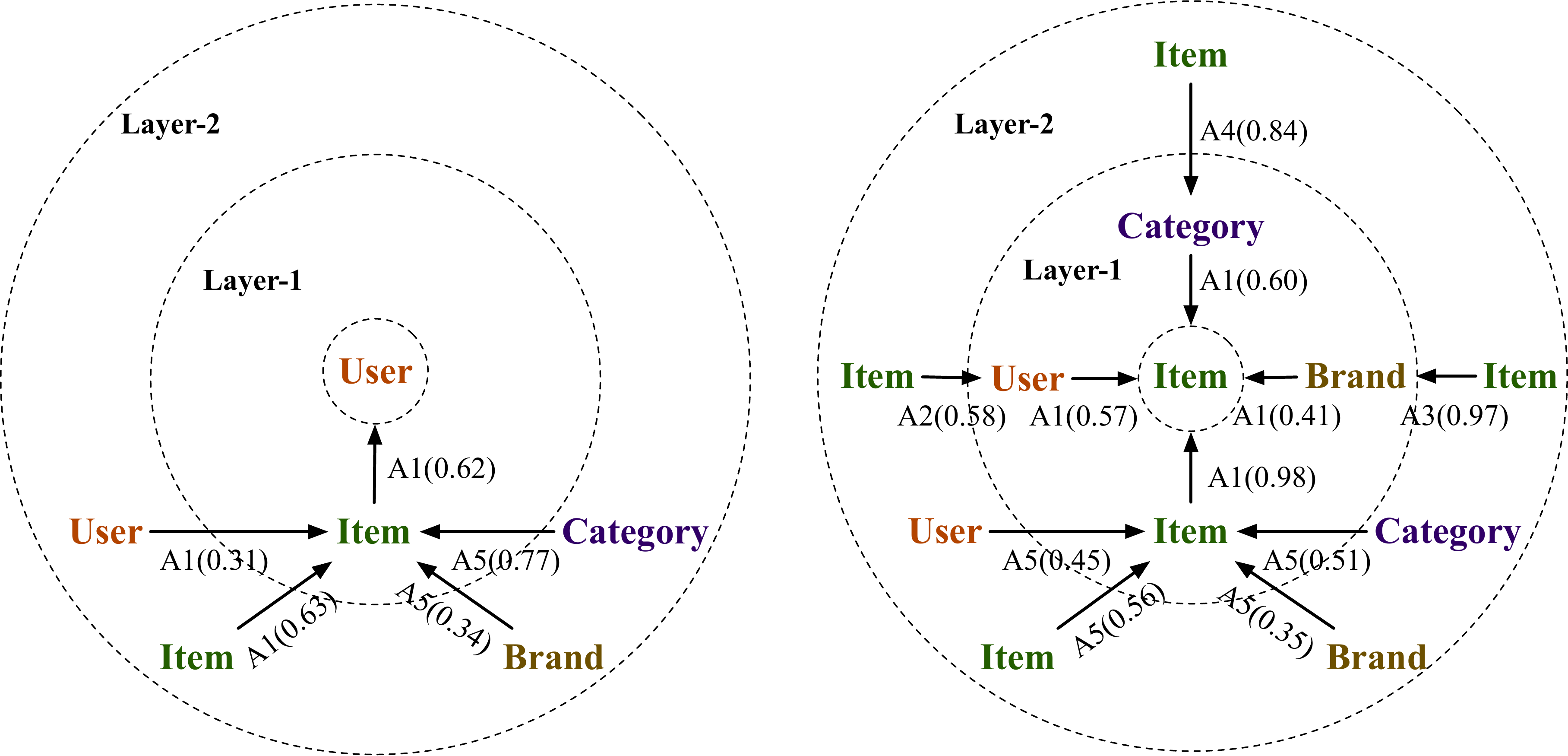}
\caption{Visualization of the learned major facet and corresponding weight of each meta relation for \textit{Amazon} data set. A1, A2, A3, A4, A5 represent 5 aspects of the relation.}
\label{fig:6}
\end{figure}

\begin{figure}
\centering
\begin{subfigure}[t]{0.49\linewidth}
\includegraphics[width=\linewidth]{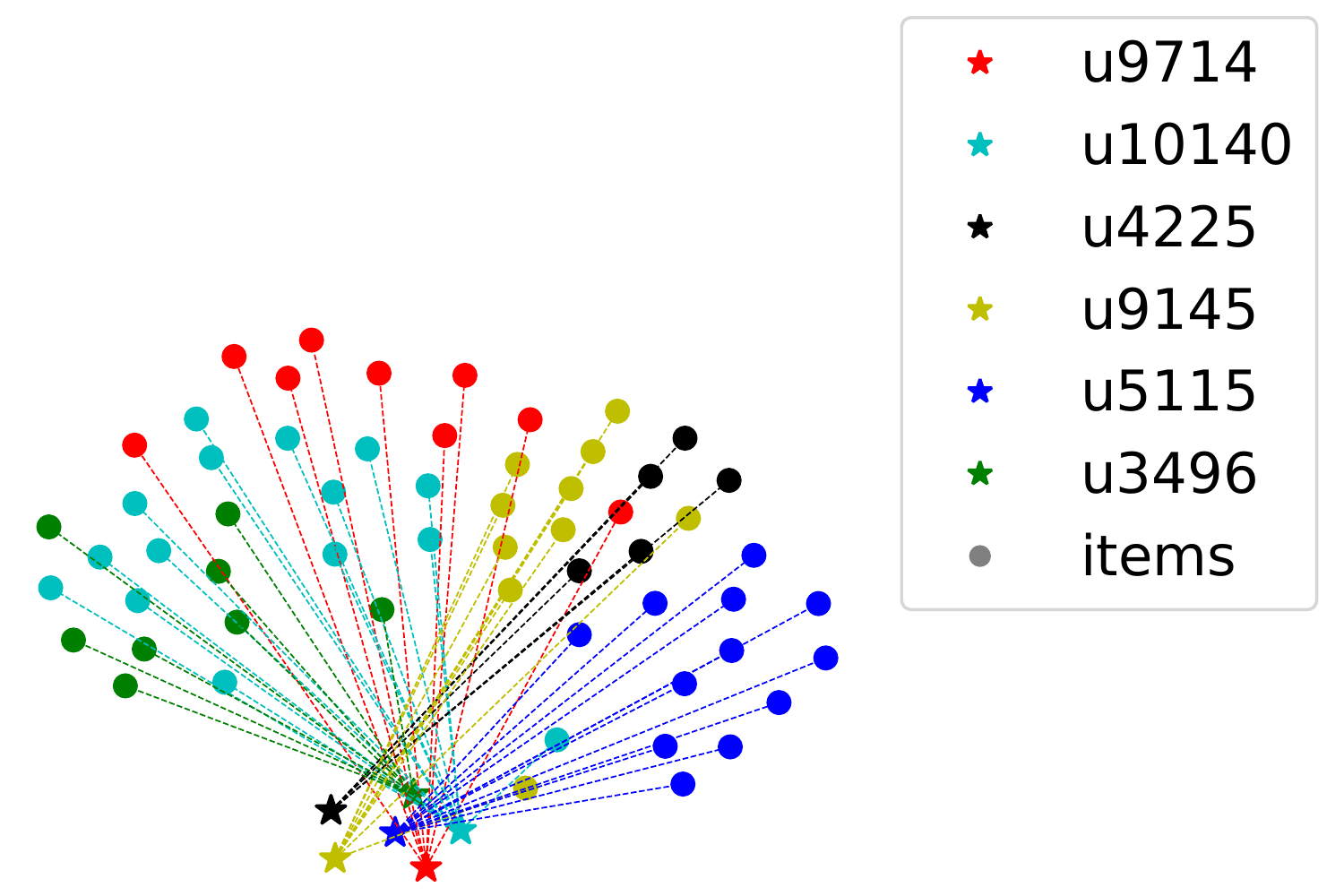}
\caption{DisenHAN} \label{fig:7_a}
\end{subfigure}
\centering
\begin{subfigure}[t]{0.49\linewidth}
\includegraphics[width=\linewidth]{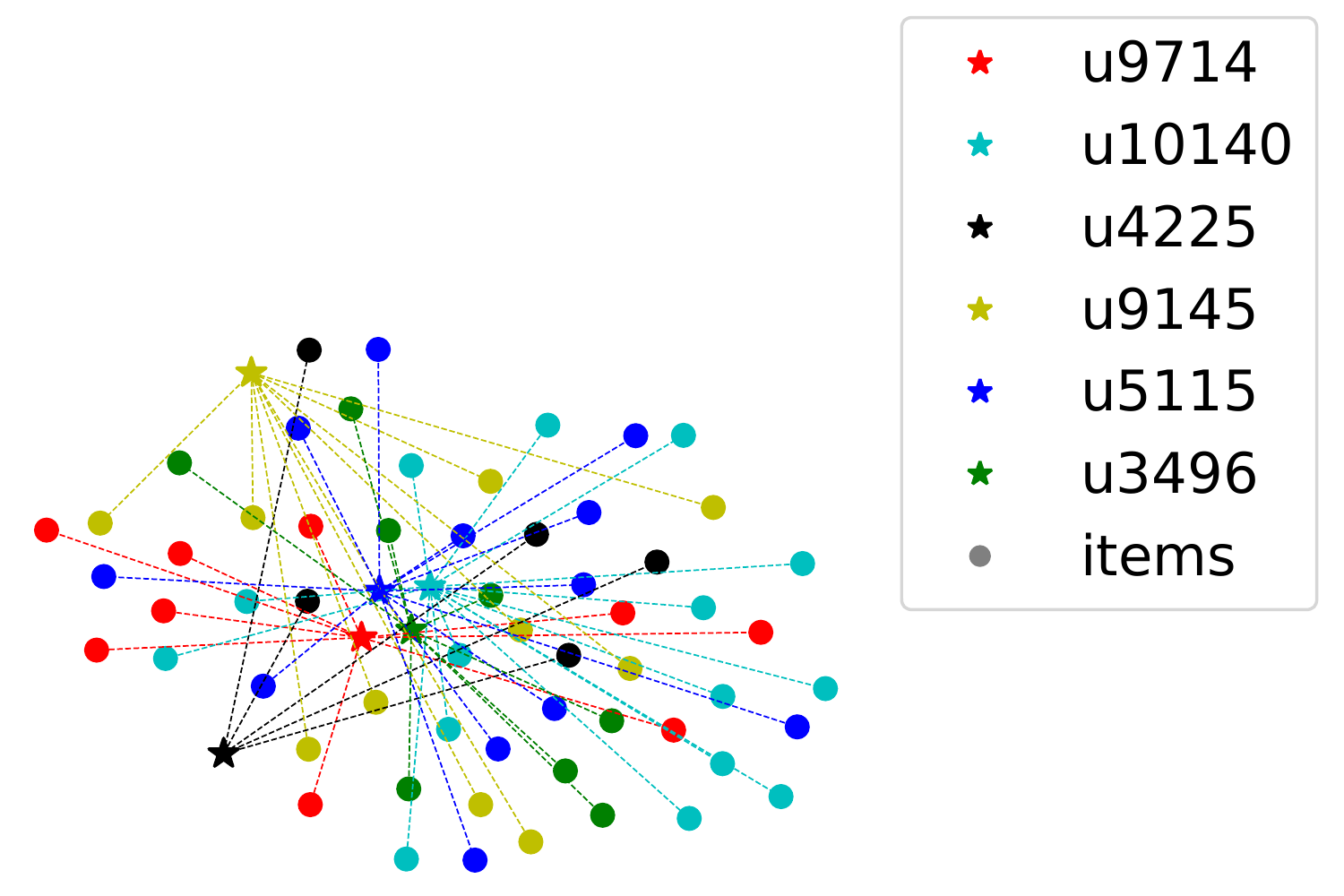}
\caption{NGCF} \label{fig:7_b}
\end{subfigure}
\caption{Visualization of the learned t-SNE transformed representations. Each star represents a user from \textit{Amazon} data set, while the points with the same color denote the relevant items.}
\label{fig:7}
\end{figure}

To further investigate how the disentangled propagation layer facilitates the embedding learning. We explore the major aspect of each meta relation in test set. Then, we visualize node embeddings of DisenHAN and NGCF to conduct a qualitative assessment of the embedding results. 
\subsubsection{Exploring Meta Relation}
To illustrate how the disentangled representation can benefit the information propagation in HIN, we pick the meta relation schema of Amazon data set and plot the average aspect weight of each meta relation. As shown in Figure~\ref{fig:6}, DisenHAN can identify the major aspect of different meta relations. For the two disentangled embedding propagation layers with 5 aspects, user-item interaction is related to aspect A1 in Layer-1. Meanwhile, since there is only one meta relation for the user in Layer-1, the other relations are all related to aspect A1 in order to calculate the matching score. While in Layer-2, aspect A2, A3, A4 are more related to item context relations which can be categorized as 
$\langle$Item, interact with$^{-1}$, User$\rangle$, $\langle$Item, has, Brand$\rangle$ and $\langle$Item, has, Category$\rangle$, respectively. This indicates that DisenHAN can capture different aspects of semantic information in HIN.  

\subsubsection{Visualization}
For a more intuitive comparison, we conduct the task of visualization. We randomly select six users from amazon data set, as well as their relevant items. Figure~\ref{fig:7_a} and Figure~\ref{fig:7_b} show the results derived from DisenHAN and  NGCF. One can quickly tell the differences that our model can effectively divide user and item nodes into two different groups. Meanwhile, the nodes correlated with different users exhibit discernible clustering, which demonstrates that different aspects of semantic information make a significant contribution to learn representation in HIN.

\section{Conclusion and Future Work}
In this work, we propose a Disentangled Heterogeneous Graph Attention Network (DisenHAN) for top-$N$ recommendation in HIN. To encode collaborative signals between users and items, we leverage embedding propagation to explicitly incorporate context information with rich semantic structure. We use meta relation to decompose high-order connectivity in HIN and propose a disentangled embedding propagation layer to aggregate different aspects of semantic information for users and items, respectively. Based on that, we can automatically generate meta paths with semantic information while capture major aspect of information flows in high-order connectivity. Extensive experiments on three real-world data sets demonstrate the effectiveness and interpretability of our model. 

For future work, we will explore whether DisenHAN is able to capture dynamic interaction of users and items in temporal graph. Besides, we also plan to extend DisenHAN to general machine learning tasks in HIN, such as node classification and node clustering tasks.

\begin{acks}
This paper is supported by National Key Research and Development Program of China with Grant No. 2018AAA0101900 / 2018AAA0101902 as well as the National Natural Science Foundation of China (NSFC Grant No. 61772039 and No. 91646202). We also thank Meituan-Dianping Group for their support. 
\end{acks}

\bibliographystyle{ACM-Reference-Format}
\bibliography{reference}










\end{document}